# Birefringence and dispersion of cylindrically polarized modes in nanobore photonic crystal fiber

T. G. Euser[1], M. A. Schmidt[1], N. Y. Joly[2,1], C. Gabriel[1,3], C. Marquardt[1,3], L. Y. Zang[1], M. Förtsch[1,3], P. Banzer[1,3], A. Brenn[1], D. Elser[1,3], M. Scharrer[2,1], G. Leuchs[1,3], and P. St.J. Russell[1,2]

[1]*Max Planck Institute for the Science of Light, Günther-Scharowsky-Str.1/Bau 24, 91058 Erlangen, Germany*

[2]*Department of Physics, University Erlangen-Nuremberg, Erlangen, Germany*

[3]*Institute for Optics, Information and Photonics,*

*University Erlangen-Nuremberg, Staudtstr. 7/B2, 91058 Erlangen*

*www.mpl.mpg.de*

We demonstrate experimentally and theoretically that a nanoscale hollow channel placed centrally in the solid glass core of a photonic crystal fiber strongly enhances the cylindrical birefringence (the modal index difference between radially and azimuthally polarized modes). Furthermore, it causes a large split in group velocity and group velocity dispersion. We show analytically that all three parameters can be varied over a wide range by tuning the diameters of the nanobore and the core.
*Prepared for arXiv.org on September 23, 2010*

## Introduction

Cylindrically polarized beams and guided modes – in which the electric field is azimuthal or radial – are finding applications in, for example, high-resolution microscopy [1], focusing of plasmons [2, 3], optical trapping [4], fiber lasers [5], quantum optics [6,7], atom optics [8, 9] and laser machining [10]. Recent papers have reported the excitation and propagation of such modes in hollow-core photonic crystal fiber (PCF) [11-13], in standard fiber with W-doping profile [14], and in all solid photonic bandgap fibers [15]. Cylindrically polarized beams can also be generated by nonlinear conversion in a few-moded photonic crystal fiber [16]. In all these cases, the radial and the azimuthal modes exhibit similar dispersion and are almost degenerate [16], allowing them to couple under the influence of bends and other perturbations.

Here we report a PCF whose sub-micron solid-glass core contains a central 180 nm wide hollow "nanobore". In such fibers there is strong electric field enhancement in the vicinity of the nanobore when the Gaussian-like linearly polarized fundamental mode is excited [17]. We show here that the nanobore has the additional effect of greatly increasing the difference in modal index, group velocity and group velocity dispersion (GVD) between radial and azimuthal modes. As a result it is possible to maintain radial and azimuthal polarization effectively against bends and perturbations, just as a conventional high-birefringence (HiBi) fiber maintains modes linearly polarized along its eigenaxes.

## Fiber

The nanobore fiber was fabricated using the conventional silica stack-and-draw procedure [18]. To avoid structural distortions during the large reduction in scale, the draw-down was carried out in three stages. The first draw resulted in a cane with an outer diameter of 1.8 mm, a core diameter of 170 µm and a hollow channel 96 µm in diameter (Fig. 1(a)). This was then drawn down to a core diameter of 31 µm containing a 16 µm hollow channel (Fig. 1(b)). During the final draw surface-tension-induced nanobore collapse was avoided by judicious use of pressure. Fig. 1(c-d) show high resolution scanning-electron-micrographs (SEMs) of the resulting fiber structure. The radii of the core and nanobore are 465 nm and 90 nm, and were found to remain constant to within ±5 nm over lengths of several meters. The air filling fraction in the cladding region was estimated from SEM images to be $F = 0.86$.

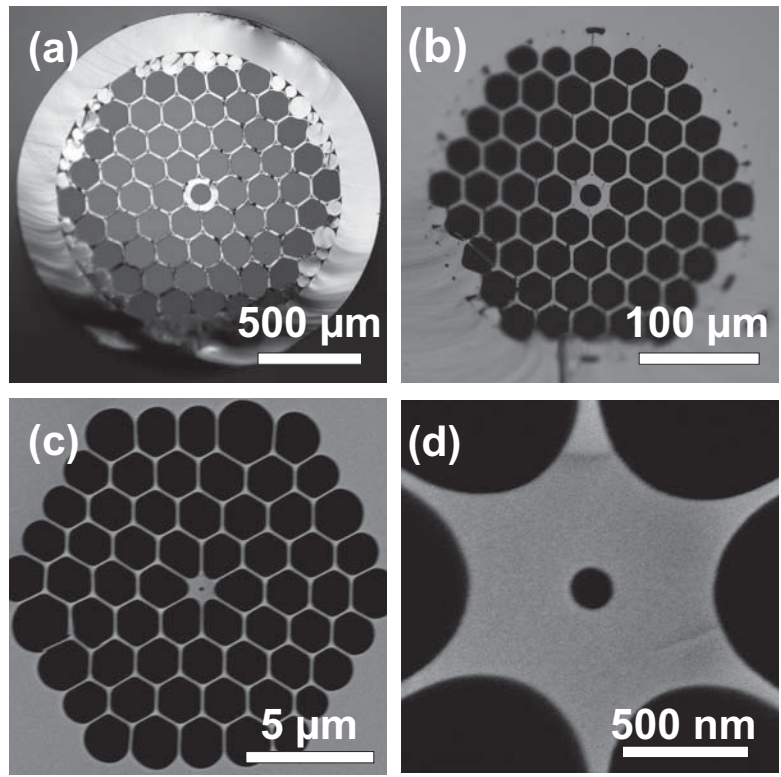

*Figure 1* Optical microscope images of (a) the first and (b) the second cane. The diameter of the hole in the core was reduced from 96 µm to 20 µm. (c) SEM of the cladding microstructure of the fused silica nanobore PCF. (d) SEM of the fiber core. The core has diameter 930 nm with a 180 nm wide nanobore located in its center.

## Set-up

The set-up for optical characterization is shown in Fig. 2(a). A tunable CW Ti:sapphire laser was used in the measurement of the near-field mode irradiance profiles (Fig. 3) and a PCF-based supercontinuum (SC) source, emitting from 480 to 1750 nm in a single transverse mode, was used for the group-delay measurements (Fig. 4) [19]. Radially and azimuthally polarized modes were generated from a linearly polarized beam using a commercial liquid-crystal-based polarization converter [20], see Fig. 2(b). A tunable liquid-crystal phase-plate (not shown) introduced a π-phase shift in half of the beam, as required to generate a doughnut mode. For each wavelength setting, the control voltage of the LC-plate was re-optimized. An additional twisted-nematic liquid crystal cell was used to switch the input polarization between vertical and horizontal, yielding an output polarization state that is either radial or azimuthal without need to change the coupling (see Fig. 2(b)). Although the orientation of the electric field varies spatially, the polarization state is locally linear, as illustrated in Fig. 2(b).

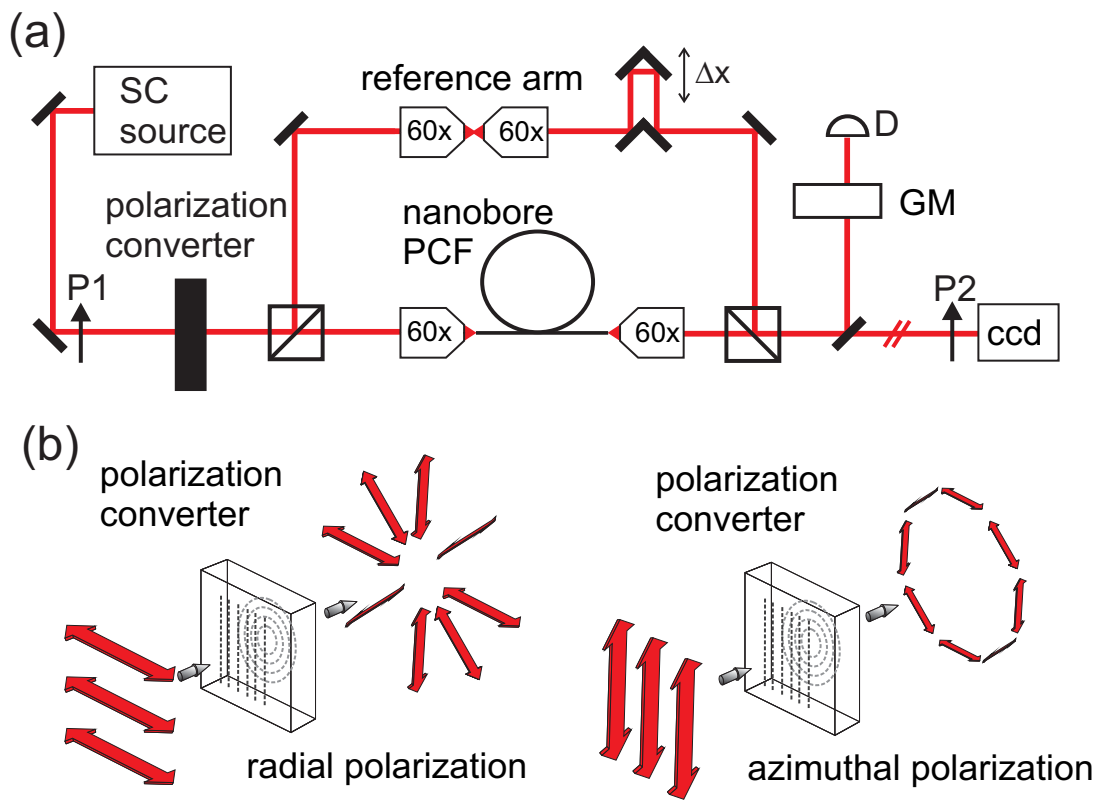

*Figure 2 (a) Experimental setup: light from a PCF-based supercontinuum (SC) source passes through a polarizer (P1) and a polarization converter [20] that converts the polarization state to either radial or azimuthal. This beam is coupled into a Mach-Zehnder interferometer. One arm includes 41 cm of the nanobore PCF shown in Fig. 1(c-d). The fiber arm and reference arm are recombined, pass a grating monochromator GM and the interference signal is measured by detector D. A CCD camera and a polarizer P2 are used to analyze the polarization state of the beam exiting the fiber. (b) Operating principle of the polarization converter using a liquid-crystal cell with alignment layers linearly rubbed on one side and circularly on the other side (from [20]).*

To measure the group delay of the modes, the cylindrically polarized beam was coupled into a Mach-Zehnder interferometer [21], one arm of which included a 41 cm long length of nanobore PCF. Using a 60× 0.85 NA objective, launch efficiencies up to 23% were obtained. The reference beam of the interferometer passed through a set of identical 60× 0.85 NA objectives (to ensure balanced dispersion between the arms) and a computer-controlled delay line (path length difference Δx), and was then recombined with the PCF signal. The combined beam was then passed, via a grating monochromator, to a photodiode and the signal monitored as a function of Δx. The short coherence length of the SC light ensures well-defined, narrow fringes (typical width 200 μm), allowing accurate measurement of the group-delay.

**Experimental results**

*Mode irradiance profiles*
Output irradiance profiles were obtained by imaging the near-field at the fiber end-face on to a CCD camera. Note that not all the sub-wavelength fine features are resolvable in the image. A doughnut-shaped irradiance distribution was observed for both modes (Fig. 3(a-b)). To analyze the polarization state, output irradiance profiles were measured for different orientations of polarizer P2 (see Fig. 3(c-f)). Some asymmetry is apparent in the profiles, which we attribute to a slightly eccentric nanobore position and core ellipticity. The data in Fig. 3(c-f) were combined to extract the spatial polarization profiles shown in Fig. 3(g-h), which match the input profiles generated by the polarization converter, demonstrating that both radially and azimuthally polarized modes are maintained after propagation in the nanobore PCF.

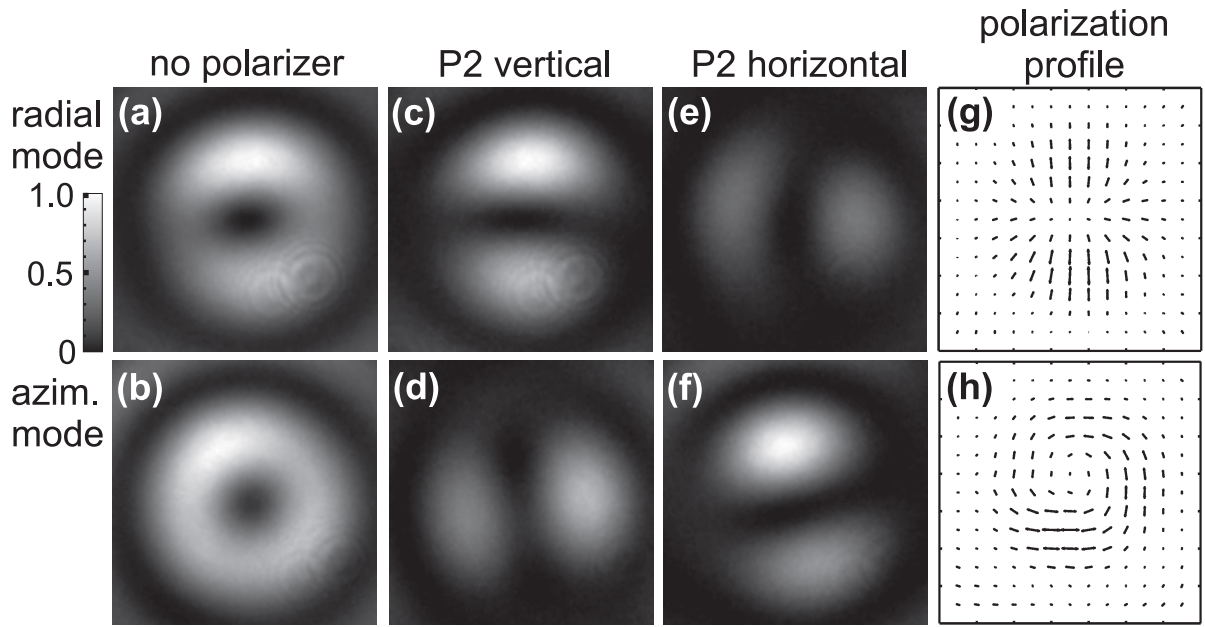

*Figure 3*: (a-f) Normalized Poynting vector profiles at λ = 820 nm for radially (top row) and azimuthally (bottom row) polarized modes obtained by imaging a 2.1×2.1 μm$^2$ region of the near-field at the fiber end-face onto a CCD camera. (c-f) Poynting vector profiles measured for orthogonal orientations of the polarizer P2 (linear grey-scale identical in each image). (g) and (h) Vector-plots of the measured electric field orientations in (a) and (b).

## Dispersion measurement

Using the set-up described above, a clean azimuthally polarized mode could be launched over the broad wavelength range 550-1060 nm, resulting in a group delay curve (Fig. 4, upper panel) that is continuous and smooth. The GVD, obtained by taking the derivative of a polynomial fit to the group delay, is plotted in the upper panel in Fig. 4. A zero dispersion wavelength (ZDW) for this mode is found at 900 nm. The radially polarized mode could only be observed in two narrower wavelength ranges, 700-860 nm and 900-1050 nm, with a ZDW at 720 nm. Its rather discontinuous group delay curve suggests the existence of two separate modes and the presence of intermodal coupling (see theory section). Interestingly, over the wavelength range 720-900 nm the azimuthally polarized mode has anomalous dispersion, whereas the radially polarized mode has normal dispersion – quite different from the case of a conventional linearly birefringent fiber, when the dispersion is very similar for both polarizations.

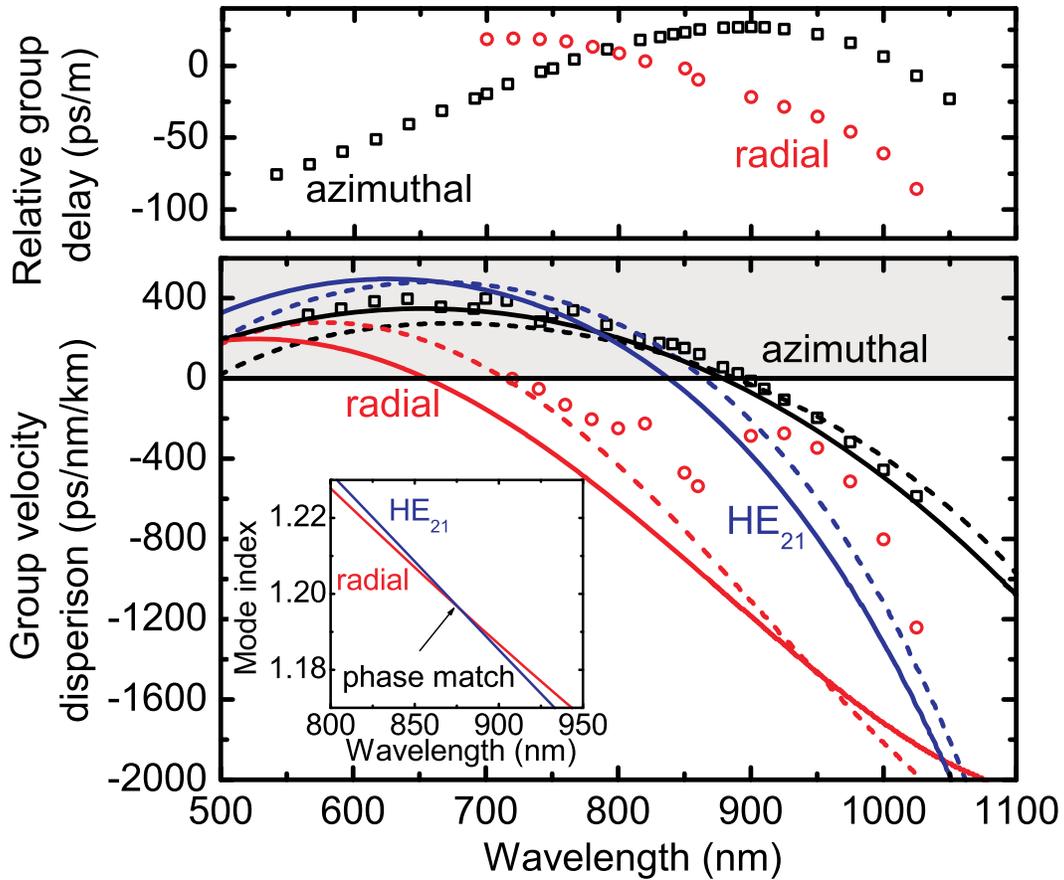

*Figure 4:* (top panel) Experimentally determined relative group delay of azimuthal and radial modes as a function of wavelength; (lower panel): measured (symbols) and calculated (curves) GVD of the azimuthal (black), radial (red) and $HE_{21}$ modes (blue). The shaded area indicates anomalous dispersion. The solid lines are the results of FE simulations, and the dashed curves are based on the analytical step-index model. (Inset): Detail of the region where the effective phase indices of the radial and $HE_{21}$ modes match.

# Theory

The optical properties of the cylindrically polarized modes were analyzed using both a quasi-analytical step-index model (based on a perfect circular strand with a nanobore at its center) and a finite-element approach based on the actual PCF structure.

## Quasi-analytical step-index model

In this model the core is approximated by a circular central hollow bore (refractive index $n_b$) and a circular annulus of fused silica (index $n_s(\lambda)$ from [22]). The refractive index of the outer medium, which in the actual fiber consists of a network of thin glass membranes, is approximated as

$$n_{cl} \approx \sqrt{F + (1-F)n_s^2} \qquad (1)$$

(the long-wavelength limit) where $F$ is the filling fraction of air.

Using the standard approach outlined in many textbooks for circular-cylindrical structures (e.g., Snyder and Love [23] or [24]), Maxwell's equations can be solved in their full vectorial form using Bessel functions, and the effective refractive indices and transverse field distributions of the guided modes calculated. The azimuthal and radial modes of the nanobore PCF are closely related to the $TE_{01}$ and $TM_{01}$ modes of a step-index fiber, with non-zero field components ($E_\varphi$, $H_r$, $H_z$) for the azimuthal and ($E_r$, $E_z$, $H_\varphi$) for the radial. As a result, the order of the relevant Bessel function for the $z$-field components is zero and the problem reduces to two 4×4 matrices, one each for the azimuthal and radial mode. Using this approach, the geometry of the structure was tuned so as to reproduce the measured GVD of the cylindrically polarized modes (Fig. 4). The best agreement was obtained for 88 nm bore radius, 540 nm core radius and $F = 0.84$, quite close to those of the PCF used in the experiments. The calculated mode irradiance distribution (not shown) as well as the spatial polarization distribution also agree well with the experimental data in Fig. 3.

The phase indices of the azimuthal and radial modes are $n_r = 1.243$ and $n_a = 1.275$, yielding a cylindrical birefringence 0.032, which translates to a beat length of 26 µm at 820 nm wavelength. In contrast, the cylindrical birefringence in a standard SMF28 fiber, designed to be single-mode at 1.55 µm but operated at 820 nm, is $4.4 \times 10^{-6}$ or four orders of magnitude smaller than in the nanobore PCF.

## Finite element calculations

Finite element calculations were performed (using COMSOL Multiphysics) on a perfectly six-fold-symmetric structure with dimensions identical to those in the actual fiber, including the Sellmeier expansion for the refractive index of silica [22]. Figure 5 shows the resulting radial and azimuthal mode Poynting vector profiles at 820 nm. Note that the Poynting vector of the radially polarized mode is discontinuous at the both inner and outer core boundaries (TM boundary conditions), whereas the azimuthal Poynting vector is continuous (TE boundary conditions). The phase indices are $n_r = 1.219$ and $n_a = 1.239$, yielding a cylindrical birefringence of 0.020, slightly smaller than predicted by the step-index model. A third mode with a phase index of 1.222, related to the $HE_{21}$ mode in standard step-index fibers, is also present (Fig. 5(c)).

The Poynting vector profiles of the modal fields, calculated after passing through a linear polarizer, are shown in Fig. 5 (d-i). The profiles compare well with the measured ones (Fig. 3(c-f)) for orthogonal orientations of the polarizer. Calculated vector electric field plots are shown in Fig. 5 (j-l), and also compare well with the measurements (Fig. 3(g-h)). Although the Poynting vector distribution and the phase index of the $HE_{21}$ mode are very similar to those of the radially polarized mode, the vector field distributions are markedly different.

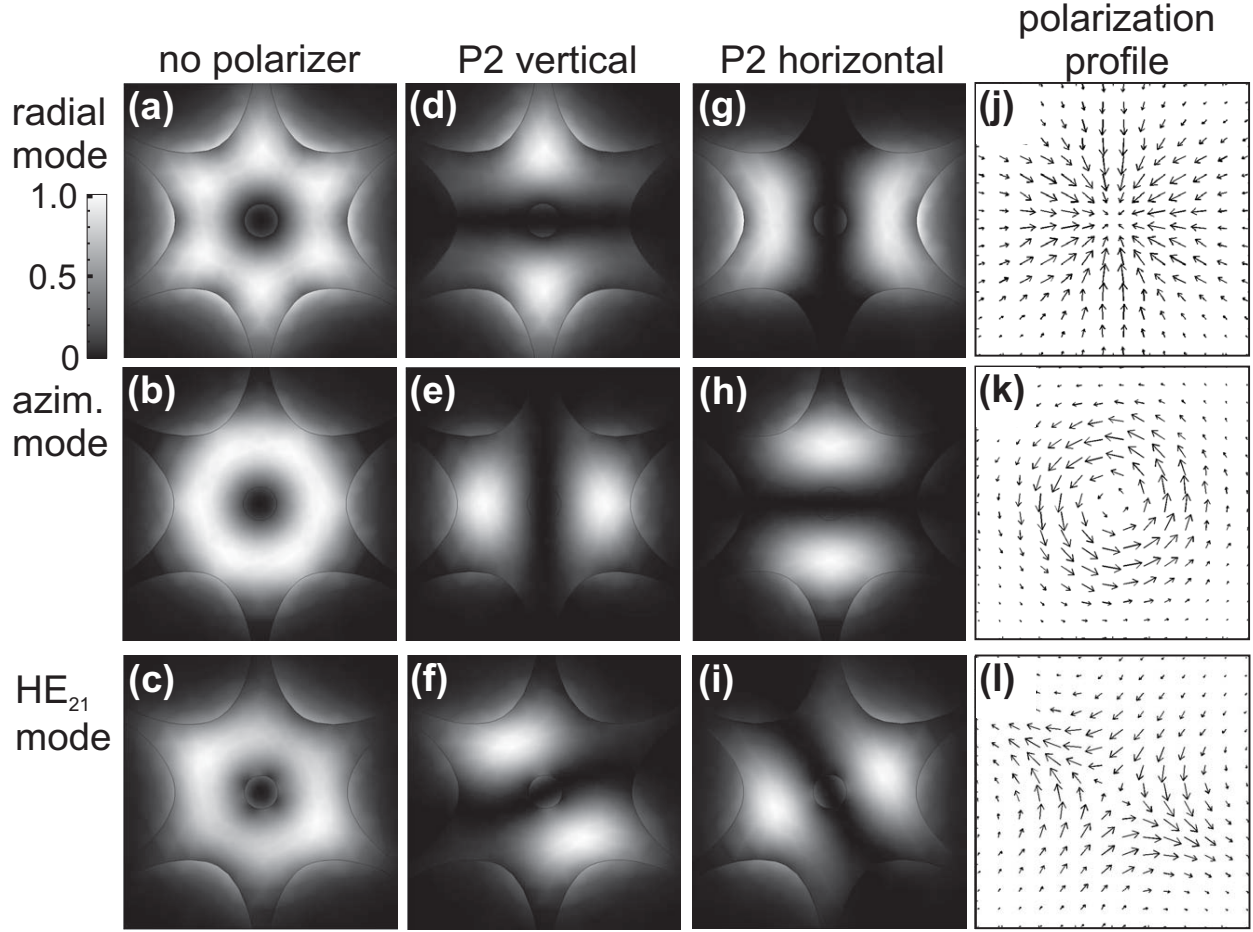

*Figure 5:* *(a-c) FE calculations of the normalized modal Poynting vector profiles at λ = 820 nm for radial (top row) and azimuthal (middle row) modes and the $HE_{21}$ higher-order mode (bottom row). All three modes show a doughnut-shaped profile. (d-i) Simulated profiles for orthogonal positions of the linear polarizer P2. (j-l) Vector electric field distributions.*

*Group velocity dispersion*
The GVD curves calculated using the FE code (solid curves) and the step index model (dashed curves) agree well with the experimental data for the azimuthal mode (Fig. 3 (b)). Reasonable agreement is also found for the radial mode in the 700-860 nm range, however, for $λ > 860$ nm theory and experiment diverge strongly, the experimental data fitting more closely to the GVD curves for the $HE_{21}$-like mode. The inset in Fig. 4(b) shows that the dispersion curves of the radial and the $HE_{21}$–like modes cross at 877 nm. Strong intermodal coupling results, which causes the measured dispersion to be a complicated mixture of that of the two modes, meaning that the radial mode dispersion cannot be cleanly measured in the range 860-900 nm.

*Tunable birefringence*

The cylindrical birefringence $B_c$ can be tuned by varying the nanobore radius $a$, as illustrated in Fig. 6, where the step-index model was used with $b = 540$ nm. Since the core radius is quite small and the outer refractive index step (using Eq. (1), this has the value 0.37 at a wavelength of 820 nm) is large, $B_c$ is significant (0.028 at 930 nm) even when $a = 0$. It is some 2.5× larger at 810 nm for a nanobore radius of 250 nm. The beat length can be as short as 11 µm and of course can be tuned to larger values by decreasing $a$. The inset in Fig. 6 shows the effective indices of the radial and azimuthal modes, calculated for $a = 0$ and 250 nm. Fig. 7 shows the dependence of $B_c$ on normalized core radius $b/\lambda$ for different ratios $a/b$. It can be seen that $B_c$ depends strongly on the ratio $a/b$.

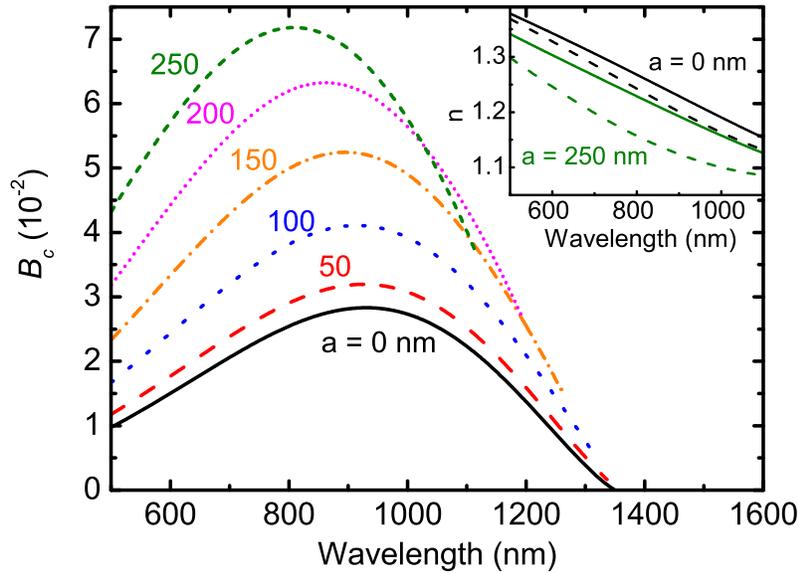

*Figure 6:* Spectral dependence of cylindrical birefringence $B_c$ for different nanobore radii a. Core-radius and air-filling fraction are 540 nm and 0.84. Inset shows the phase indices of the azimuthal (solid lines) and radial (dashed lines) modes for a = 0 (black) and 250 nm (green).

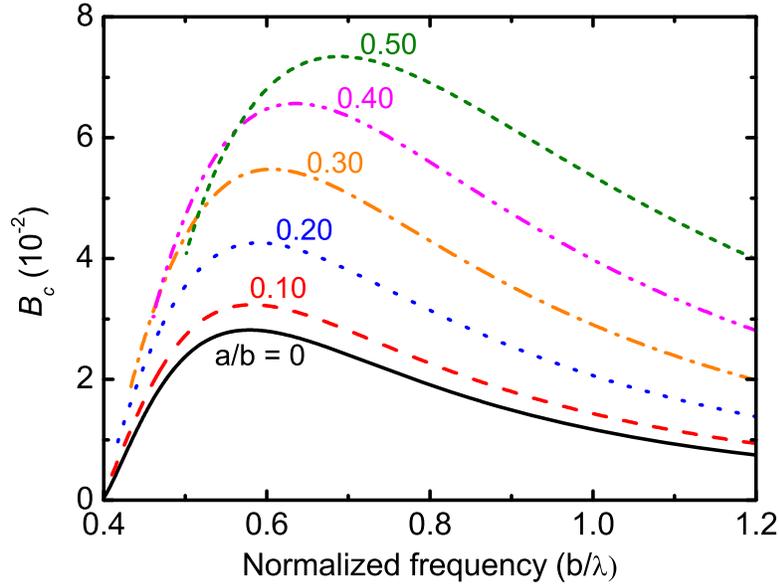

*Figure 7:* Calculated cylindrical birefringence versus normalized frequency for different values of the ratio a/b (step-index model). Core-radius and air-filling fraction are 540 nm and 0.84. The silica refractive index is kept constant at $n_s$ = 1.45. The low frequency edge of each curve (near $b/\lambda$ = 0.4) corresponds to the cut-off frequency of the radial mode.

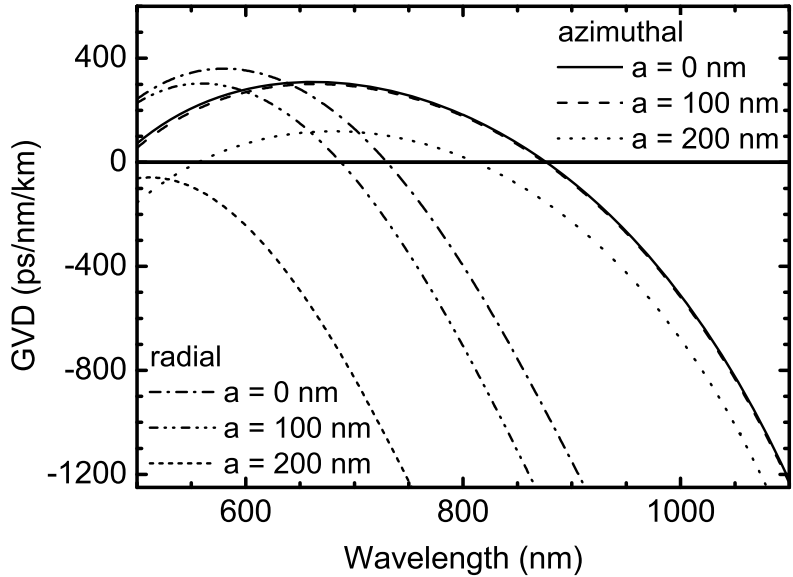

*Figure 8*: Calculated GVD curves for radial and azimuthal modes (step-index model). Core-radius and air-filling fraction are 540 nm and 0.84. Data are plotted for three different values of a.

*Tunable dispersion*

The ZDWs move to shorter wavelength as *a* increases, as seen in Fig. 8 for three PCFs with $b = 540$ nm and nanobore radii $a = 0$, 100 and 200 nm. The radial GVD is more strongly affected by the presence of the nanobore than the azimuthal. For $a > 195$ nm the radial GVD remains normal for all wavelengths (i.e., there is no ZDW). The great difference in GVD tunability between the azimuthal and radial modes means that the positions of their ZDWs can be almost independently controlled (see Fig. 9), which could be of interest in nonlinear applications.

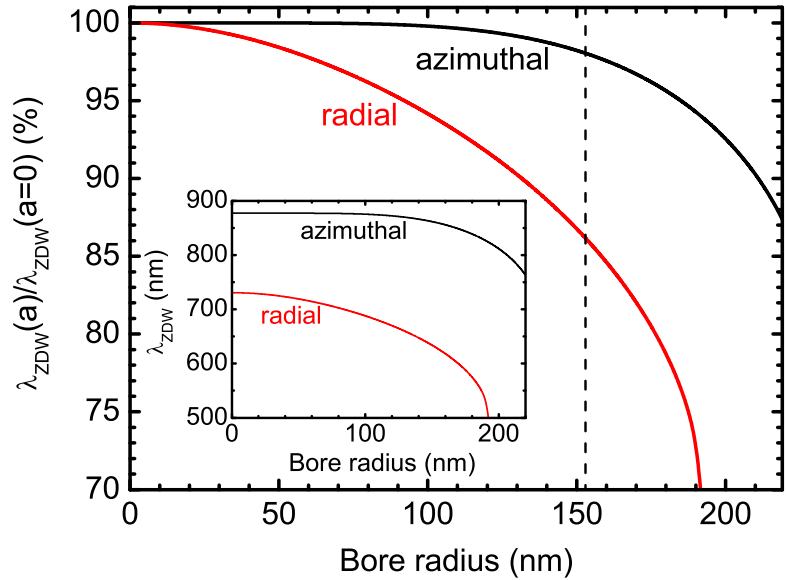

*Figure 9:* Calculated relative change of the second (long wavelength) zero dispersion wavelength ($\lambda_{ZDW}$) as function of nanobore radius for the azimuthal (black) and radial (red) modes (step-index model). Curves have been normalized to the value of $\lambda_{ZDW}$ at $a = 0$. Inset shows the absolute values of $\lambda_{ZDW}$. Core-radius and air-filling fraction are 540 nm and 0.84. For $a < 153$ nm (dashed line), $\lambda_{ZDW}$ for the azimuthal mode decreases by a maximum of 2% (~17 nm), while $\lambda_{ZDW}$ for the radial mode decreases by a maximum of 14% (~100 nm).

## Conclusions and perspective

Nanobore PCF effectively maintains azimuthal and radial polarization states. In addition, the phase indices, group velocities and GVD of the orthogonal azimuthal and radial modes are strongly different and can be widely tuned by varying the core and nanobore radii. A simple quasi-analytical step-index model provides convenient approximate solutions for the modal indices and field profiles, which compare well with FE simulations of the actual fiber structure. These unique characteristics are interesting for many applications, for example, the azimuthal mode in the range of anomalous dispersion has been used for nonlinear quantum-squeezing, permitting for the first time creation of a continuous-variable hybrid-entangled state [7]. The nanobore can be filled with other materials, such as metals or nonlinear glasses, e.g., tellurite [25, 26]. Since the nanobore fiber is polarization maintaining over a broad wavelength range, it could be used to transmit broad-band radially-polarized modes with high fidelity along extended curved paths, with applications in, e.g., high resolution microscopy systems [27].

## Acknowledgements

We thank Silke Rammler for help in fabricating the fiber, Namvar Jahanmer for the scanning electron micrographs, and Juan Carlos Loredo Rosillo for help with CAD drawings used for FEM simulations.